\documentclass[twocolumn,pre,showpacs,preprintnumbers,amsmath,amssymb]{revtex4}

\usepackage[dvips]{graphicx}% Include figure files
\usepackage{dcolumn}% Align table columns on decimal point
\usepackage{bm}% bold math

\begin{document}

\title{Wavelet analysis of epileptic spikes}

\author{Miroslaw Latka}
\email{mirek@if.pwr.wroc.pl}
\author{Ziemowit Was}
\email{zwas@wp.pl}
\affiliation{
Institute of Physics, Wroclaw University of Technology, Wybrzeze Wyspianskiego 27,
                    50-370 Wroclaw, Poland\\
}%

\author{Andrzej Kozik}
\email{andrzejkozik1@wp.pl}
\affiliation{
Video EEG Lab, Department of Child Neurology, \\
Regional Medical Center,
ul. Traugutta 116, 40-529 Wroclaw, Poland
}\email{andrzejkozik1@wp.pl}

\author{Bruce J. West}
\email{westb@aro-emh1.army.mil}
\affiliation{Mathematics Division, Army Research Office,
             P.O. Box 12211, Research Triangle, NC 27709-2211, USA
}

\date{January 17, 2003}

\begin{abstract}

Interictal spikes and sharp waves in human EEG are characteristic
signatures of epilepsy. These potentials originate as a result of
synchronous, pathological discharge of many neurons. The reliable
detection of such potentials has been the long standing problem in
EEG analysis, especially after long-term monitoring became common
in investigation of epileptic patients. The traditional definition
of a spike is based on its amplitude, duration, sharpness, and
emergence from its background. However, spike detection systems
built solely around this definition are not reliable due to the
presence of numerous transients and artifacts. We use wavelet
transform to analyze the properties of EEG manifestations of
epilepsy. We demonstrate that the behavior of wavelet transform of
epileptic spikes across scales can constitute the foundation of a
relatively simple yet effective detection algorithm.

\end{abstract}

\maketitle

Recordings of human brain electrical activity (EEG) have been the
fundamental tool for studying the dynamics of cortical neurons
since 1929. Even though the gist of this technique has essentially
remained the same, the methods of EEG data analysis have
profoundly evolved during the last two decades. In 1985 Babloyantz
\textit{et al.} demonstrated that certain nonlinear measures,
first introduced in the context of chaotic dynamical systems,
changed during slow-wave sleep \cite{babloyantz85}. The flurry of
research work that followed this discovery focused on the
application of nonlinear dynamics in quantifying brain electrical
activity during different mental states, sleep stages, and under
the influence of the epileptic process (for a review see for
example \cite{bwest95,pritchard92}). It must be emphasized that a
straightforward interpretation of neural dynamics in terms of such
nonlinear measures as the largest Lyapunov exponent or the
correlation dimension is not possible since most biological time
series, such as EEG, are nonstationary and consequently do not
satisfy the assumptions of the underlying theory. On the other
hand, traditional power spectral methods are also based on quite
restrictive assumptions but nevertheless have turned out to be
successful in some areas of EEG analysis. Despite these technical
difficulties, the number of applications of nonlinear time series
analysis  has been growing steadily and now includes the
characterization of encephalopaties \cite{stam99}, monitoring of
anesthesia depth \cite{widman00}, characteristics of seizure
activity \cite{casdagli97}, and prediction of epileptic seizures
\cite{lehnertz98}. Several other approaches are also used to
elucidate the nature of electrical activity of the human brain
ranging from coherence measures \cite{nunez97,nunez99} and methods of
nonequilibrium statistical mechanics \cite{ingber82} to
complexity measures \cite{rezek98, putten01}.

One of the most important challenges of EEG analysis has been the
quantification of the manifestations of epilepsy. The main goal is
to establish a correlation between the EEG and clinical or
pharmacological conditions. One of the possible approaches is
based on the  properties of the interictal EEG (electrical
activity measured between seizures) which  typically consists of
linear stochastic background fluctuations interspersed with
transient nonlinear spikes or sharp waves. These transient
potentials originate as a result of a simultaneous pathological
discharge of neurons within a volume of at least several mm$^3$.

The traditional definition of a spike is based on its amplitude,
duration, sharpness and  emergence from its background
\cite{gotman76,gotman82}. However, automatic epileptic spike
detection systems based on this direct approach suffer from false
detections in the presence of numerous types of artifacts and
non-epileptic transients. This shortcoming is particularly acute
for long-term EEG monitoring of epileptic patients which became
common in 1980s. To reduce false detections Gotman and Wang
\cite{gotman91} made the process of spike identification dependent
upon the state of EEG (active wakefulness, quiet wakefulness,
desynchronized EEG, phasic EEG, and slow EEG). This modification
leads to significant overall improvement provided that state
classification is correct.

Diambra and Malta \cite{diambra99} adopted nonlinear prediction
for epileptic spike detection. They demonstrated that when  the
model's parameters are adjusted during the \textquotedblleft
learning\textquotedblright  phase to assure good predictive
performance for stochastic background fluctuations, the appearance
of an interictal spike is marked by a very large forecasting
error. This novel approach is appealing because it makes use of
changes in EEG dynamics. One expects good nonlinear predictive
performance when the dynamics of the EEG interval used for
building up the model is similar to the dynamics of the interval
used for testing. However, it is uncertain at this point whether
it is possible to develop a robust spike detection algorithm based
solely on this idea.

As Clark \textit{et al.} put it succinctly, automatic EEG analysis
is a formidable task because of  the lack of \textquotedblleft
...features that reflect the relevant information
\textquotedblright \cite{clark95}. Another difficulty is the
nonstationary nature of the spikes and the background in which
they are embedded. One technique developed for the treatment of
such nonstationary time series is wavelet analysis
\cite{mallat98,unser96}. The goal of this paper is to characterize
the epileptic  spikes and sharp waves in terms of the properties
of their wavelet transforms. In particular, we search for features
which could be important in the detection of epileptic events.

The wavelet transform  is an  integral transform for which the set
of basis functions, known as wavelets, are well localized both in
time and frequency. Moreover, the wavelet basis can be constructed
from a single function $\psi(t)$ by means of translation and
dilation:
\begin{equation}
\psi_{a;t_0}=\psi \left(\frac{t-t_0}{a}\right).
\label{wbasis}
\end{equation}
$\psi(t)$ is commonly referred to as the mother function or analyzing
wavelet.
The wavelet transform of function $h(t)$ is defined as
\begin{equation}
W(a,t_0)=\frac{1}{\sqrt a}\int_{-\infty }^{\infty }h(t)\psi_{a;t_0}^{*}dt,
\label{waveletDef}
\end{equation}
where $\psi^{*}(t)$ denotes the complex conjugate of $\psi(t)$.
The continuous wavelet transform of a discrete time series
$\{h_i\}^{N-1}_{i=0}$ of length $N$ and equal spacing $\delta t$
is defined as
\begin{equation}
W_n(a)=\sqrt{\frac{\delta t}{a}}\sum^{N-1}_{n'=0}
       { h_{n'}\psi ^{*}\left [\frac{(n'-n)\delta t}{a}\right ]}.
\label{dwt}
\end{equation}

The above convolution can be evaluated for any of $N$  values of the time index $n$.
However, by choosing all  $N$ successive time index values, the convolution theorem allows us
to calculate all $N$ convolutions simultaneously in Fourier space using
a discrete Fourier transform (DFT). The DFT of $\{h_i\}^{N-1}_{i=0}$ is
\begin{equation}
\hat{h}_k=\frac{1}{N}\sum_{n=0}^{N-1}{h_n e^{-2 \pi i k n/N}},
\label{dft}
\end{equation}
where $k=0,...,N-1$ is the frequency index. If one notes that the
Fourier transform of a function $\psi(t/a)$ is $|a|\hat{\psi}(a
f)$ then by the convolution theorem
\begin{equation}
W_n(a)=\sqrt{a \delta t}\sum^{N-1}_{k=0}{\hat{h}_n
       \psi^{*}(a  f_k) e^{2 \pi i f_k n \delta t }},
\label{fourierrep}
\end{equation}
frequencies $f_k$ are  defined in the conventional way.
Using (\ref{fourierrep}) and a standard fast Fourier transform (FFT)
routine it is possible to efficiently  calculate the continuous wavelet transform
(for a given scale $a$) at all $n$ simultaneously \cite{torrence98}.
It should be emphasized that formally equation (\ref{fourierrep}) does not
yield the discrete linear convolution corresponding to (\ref{dwt}) but rather 
a discrete circular convolution in which the shift $n'-n$ is taken modulo $N$.
However, in the context of this work, this problem does not give rise
to any numerical difficulties. This is because,
for purely practical reasons, the  beginning and the end of the analyzed part
of data stream are not taken into account during the EEG spike detection.

From a plethora of available mother wavelets, we employ the Mexican hat
\begin{equation}
\psi(t)= \frac{2}{\sqrt 3}\pi^{-1/4}(1-t^2)e^{-t^2/2}
\label{mhpsi}
\end{equation}
which is particularly suitable for studying epileptic events.

\begin{figure}[]
\includegraphics[height=3.0in]{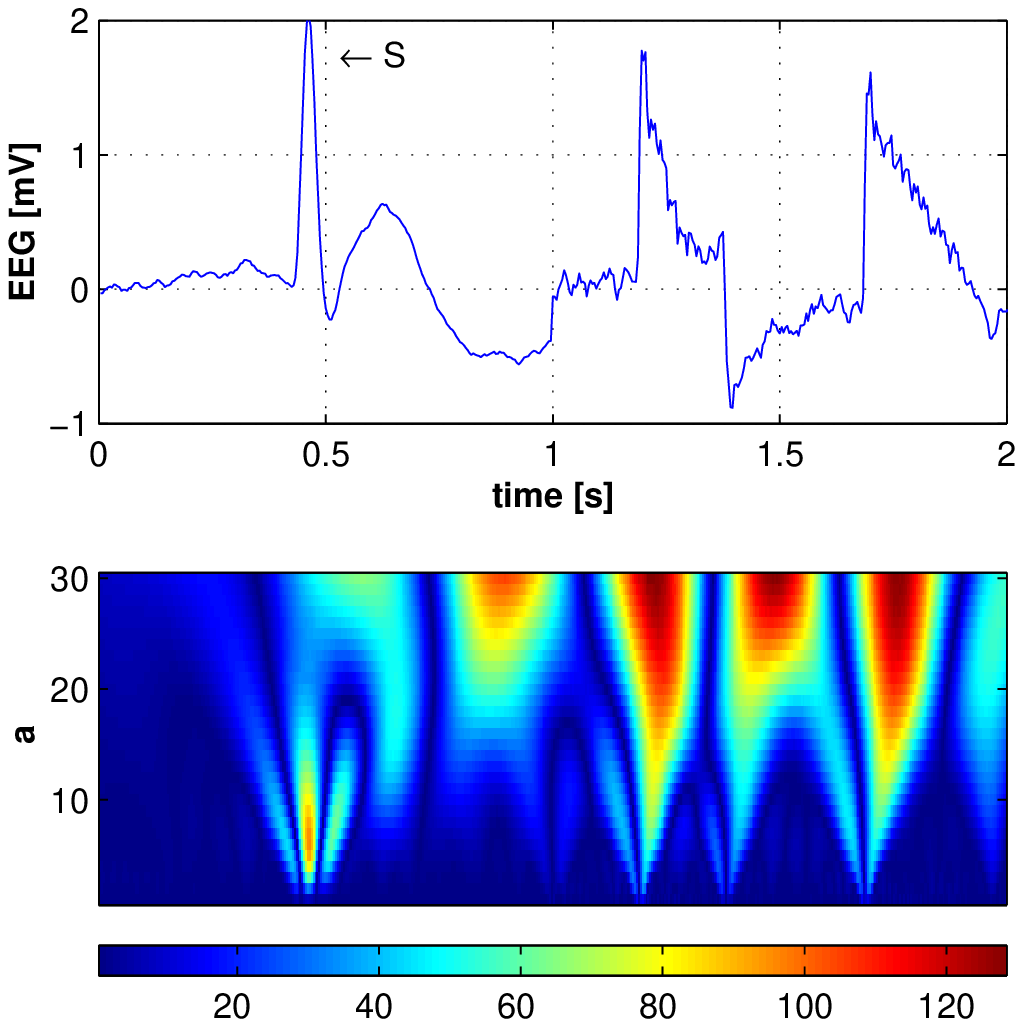}
\caption{Top panel: simple epileptic spike (marked by S) followed
by two artifacts. Bottom panel:  contour map of the absolute value
of the Mexican hat wavelet coefficients (arbitrary units)
calculated for the EEG signal shown above. The shades of blue
correspond to low values and the shades of red to high values.}
\label{WaveletMap}
\end{figure}

In the top panel of Fig. \ref{WaveletMap} we present two pieces of
the EEG recording joined at approximately $t=1s$. The digital 19
channel recording sampled at 240 Hz was obtained from a juvenile
epileptic patient according to the international 10-20 standard
with the reference average electrode. The epileptic spike in this
figure (marked by the arrow) is followed by two artifacts. The
bottom panel of Fig. \ref{WaveletMap} displays the contour map of
the absolute value of Mexican hat wavelet coefficients $W(a,t_0)$.
It is apparent that the red  prominent ridges correspond to the
position of either spike or the motion artifacts. What is most
important, for small scales, $a$, the values of the wavelet
coefficients for the spike's ridge are much larger than those for
the artifacts. The peak value along the spike ridge corresponds to
$a=7$. In sharp contrast, for the range of scales used in Fig.
\ref{WaveletMap} the absolute value of coefficients $W(a,t_0)$ for
the artifacts grow monotonically with $a$.

\begin{figure}[]
\includegraphics[height=3.0in]{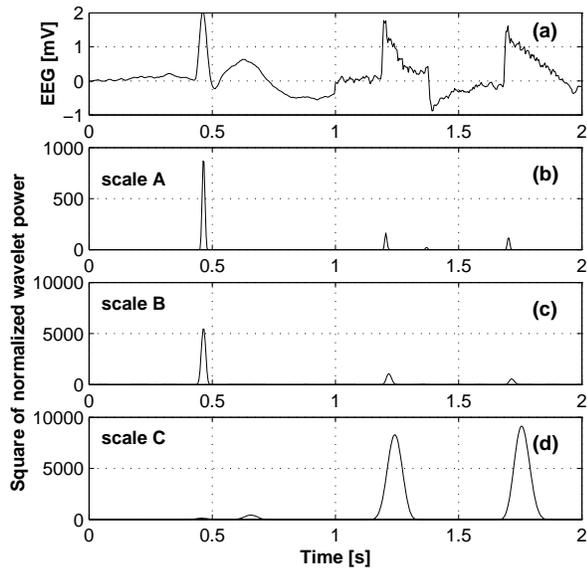}
\caption{Square of normalized wavelet power for three different scales
        $A<B<C$ (Panels (b)-(d)).
        The EEG signal shown in panel (a) is the same  as the one used in Fig. \ref{WaveletMap}
}

\label{wpS}
\end{figure}

The question arises as to whether
the behavior of the wavelet transform as a function of scale can be used
to develop a reliable detection algorithm. The first step in this direction
is to use the normalized wavelet power
\begin{equation}
 w(a,t_0)=W^2(a,t_0)/\sigma^2
\label{wpower}
\end{equation}
instead of the wavelet coefficients to reduce the dependence on
the amplitude of the EEG recording. In the above formula
$\sigma^2$ is the variance of the portion of the signal being
analyzed (typically we use pieces of length 1024 for EEG tracings
sampled at 240 Hz). In actual numerical calculations we prefer to
use the square of $w(a,t_0)$ to merely increase the range of
values  analyzed during the spike detection process. In Fig.
\ref{wpS} $w^2$ for the signal used in Fig. \ref{WaveletMap} is
plotted for three scales $A=3$, $B=7$ and $C=20$.

In the most straightforward approach, we identify
an EEG transient potential as a simple or isolated epileptic spike if and only if:

$\bullet$ the value of $w^2$ at $a=7$ is greater than a
predetermined threshold value $T_1$,

$\bullet$ \textit{the square of normalized wavelet power decreases from scale $a=7$ to $a=20$},

$\bullet$ the value of $w^2$ at $a=3$ is greater than a
predetermined threshold value $T_2$.

The threshold values $T_1$
and $T_2$ may be considered as the model's parameters which can be
adjusted to achieve the desired \textit{sensitivity} (the ratio of
detected epileptic events to the total number of epileptic events
present in the analyzed EEG tracing) and \textit{selectivity} (the
ratio of epileptic events to the total number of events marked by
the algorithm as epileptic spikes).

\begin{figure}[]
\includegraphics[height=3.0in]{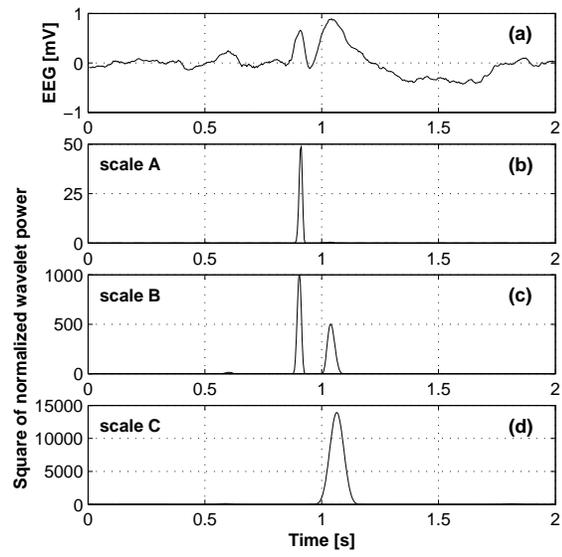}
\caption{ (a) Epileptic spike - slow wave complex. The amplitude
         of the slow wave is comparable to that of the spike.
         The square of normalized wavelet power for this signal
         is shown in panels (b)-(d) for three different scales $A<B<C$.}
\label{wpSW}
\end{figure}

While this simple algorithm is
quite effective for simple spikes such as one shown in Fig.
\ref{WaveletMap} it fails for the common case of an epileptic
spike accompanied by a slow wave with comparable amplitude. The
example of such complex is given in Fig. \ref{wpSW}(a). The
overlap of the negative tail of the Mexican hat with the slow wave
yields the  inherently low  values of $w^2$ at scale $A$  (panel
(b)) and scale $B$ (panel (c)) as compared to those characteristic
of the \textquotedblleft  isolated\textquotedblright  spike.
Nevertheless, the normalized wavelet power does decrease from
scale $B$ to $C$. Consequently, in the same vein  as the argument
we presented above, we can develop an algorithm which detects the
epileptic spike in the vicinity of a slow wave by calculating the
following linear combination of wavelet transforms:
\begin{equation}
\tilde{W}(a,t_0)= c_1 W(a,t_0) + c_2 W(a_s,t_0+ \tau)
\label{complex}
\end{equation}
and checking whether the square of corresponding normalized power
$\tilde{w}(a,t_0)=\tilde{W}^2(a,t_0)/\sigma ^2$at scales $a=3$ and
$a=7$ exceeds the threshold value $\tilde{T}_1$ and $\tilde{T}_2$,
respectively. The second term in (\ref{complex}) allows us to
detect the slow wave which follows the spike. The parameters $a_s$
and $\tau$ are chosen to maximize the overlap of the wavelet with
the slow wave. For the Mexican hat we use $a_s=28$ and $\tau=0.125
s$. By varying the values of coefficients $c_1$ and $c_2$, it is possible to
control the relative contribution of the spike and the slow wave to
the linear combination (\ref{complex}).

For testing purposes, we built up the database of artifacts  and
spikes. We made available some of these EEG tracings at
\cite{homepage} along with the examples of the numerical
calculations. While the analysis of the pieces of EEG recordings such as
those shown in Fig. \ref{wpS} and \ref{wpSW} is essential in determining
the generic properties of epileptic events, it can hardly reflect the difficulties
one can encounter in interpretation of clinical EEG. Therefore we selected
four \textit{challenging} EEG tracings with 340 epileptic events. The algorithm
described in this work marked 356 events out of which 239 turned out to be the epileptic events.
Thus the sensitivity of the algorithm was 70\% and its selectivity was equal to 67\%.
We then analyzed the same tracings with the leading commercial spike detector
developed by the Persyst Development Corporation (Insight 2001.07.12).
This software marked 654 events out of which 268 were epileptic events. Thus slightly  better
sensitivity of 79\% was achieved at the expense of  the low 41\% selectivity.
The performance of preliminary numerical implementation of the detection algorithm presented  in this
work is excellent and allows to process 24 hour EEG recording (19 channels sampled at 240 Hz)
in a matter of minutes on the average personal computer.

The goal of wavelet analysis of
the two types of spikes, presented in this paper, was to elucidate
the approach to epileptic events detection which explicitly hinges on the
behavior of  wavelet power spectrum of EEG signal \textit{across}
scales and not merely on its values. Thus, this approach
is  distinct not only from the detection algorithms based upon  discrete
multiresolution representations of EEG recordings
\cite{blanco96,dattellis97,sartoretto99,calvagno00,gutierrez01} but also
from the method developed by Senhadji and Wendling  which employs
continuous wavelet transform \cite{senhadji02}.

Epilepsy is a common disease which affects 1-2\% of the population and about 4\% of children
\cite{jallon02}.  In some epilepsy syndromes interictal paroxysmal discharges of cerebral neurons
reflect the severity of the epileptic disorder and  themselves are believed to contribute
to the progressive disturbances in cerebral functions (\textit{eg.} speech impairment,
behavioral disturbances) \cite{engel01}. In such cases  precise quantitative spike
analysis   would be extremely important. The epileptic event detector  described in this paper 
was developed with this particular goal in mind and its application to the studies of
the Landau-Kleffner syndrome will be presented elsewhere.

\bibliography{wavelet2}

\end{document}